\documentclass[authoryear,1p,review,doublespacing,12pt]{elsarticle}
\usepackage{amssymb}
\usepackage{amsmath}
\usepackage{wasysym}
\usepackage{natbib}
\usepackage{graphicx}
\usepackage{xcolor}
\usepackage[displaymath]{lineno}
\usepackage[pdftitle={Yongxiang Huang: JHydrolb},bookmarks=true,citecolor=red,colorlinks=false]{hyperref}
 \linenumbers

 \usepackage{ulem}

\renewcommand{\eqref}[1]{~(\ref{#1})}

\newcommand{\red}[1]{\textcolor{black}{#1}}

\journal{Journal of Marine Systems}

\begin{document}

\begin{frontmatter}


\title{Time dependent intrinsic correlation analysis of temperature and dissolved oxygen time series using empirical mode decomposition  }

\author[China]{Yongxiang Huang}
\ead{yongxianghuang@gmail.com}
\author[France]{Fran\c{c}ois  G. Schmitt\corref{corauthF}}
\ead{francois.schmitt@univ-lille1.fr}


\address[China]{Shanghai Institute of Applied Mathematics and Mechanics, Shanghai Key Laboratory of Mechanics in Energy Engineering, Shanghai University,
Shanghai 200072, People\rq{}s Republic of China}

\address[France]{CNRS and University of Lille 1, Laboratory of Oceanology and Geosciences, UMR 8187 LOG,  62930 Wimereux, France}

\begin{abstract}

In the marine environment, many fields have fluctuations over a large range of different spatial and temporal scales. These quantities can be nonlinear \red{and} non-stationary, and often interact with each other. A good method to study the multiple scale dynamics of such time series, and their correlations, is needed. In this paper an application of an empirical mode decomposition based time dependent intrinsic correlation, \red{of} two coastal oceanic time series, temperature and dissolved oxygen (saturation percentage) is presented. The two time series are recorded every 20 minutes \red{for} 7 years, from 2004 to 2011. The application of the Empirical Mode Decomposition on such time series is illustrated, and  the power spectra of the time series are estimated using the Hilbert transform (Hilbert spectral analysis). Power-law regimes are found with slopes of 1.33 for dissolved oxygen and 1.68 for temperature at high frequencies (between 1.2 and 12 hours) \red{with} both close to 1.9 for lower frequencies (time scales from 2 to 100 days). Moreover, the time evolution and scale dependence of cross correlations between both series are considered. 
The trends are perfectly anti-correlated. The modes of mean year 3  and 1 year have also negative correlation, whereas higher frequency modes have a much smaller correlation. The estimation of time-dependent
intrinsic correlations helps to show patterns of correlations at different scales, for different modes.
\end{abstract}


\end{frontmatter}

\pagebreak

\section{Introduction}

\red{Generally} in geosciences, \red{but} especially in the marine environment, many fields have fluctuations over a large range of spatial and temporal scales. To study their dynamics and estimate their variations at all scales, high frequency measurements are needed \citep{Dickey1991,Chavez1997,Chang2001}. Here  a time series 
obtained from automatic measurements in a moored \red{buoy} station in coastal waters of Boulogne-sur-mer (eastern English Channel, France) is considered, recorded every 20 minutes from 2004 to 2011 \citep{Zongo2011a,Zongo2011b}. This fixed buoy station  \red{can} record \red{various} biogeochemical parameters simultaneously. Here,  mainly to illustrate the application of 
a new method for multi-scale data analysis there is a focus on two parameters: temperature, due to its obvious importance, influenced by the dynamics, by meteorology, and as a link with ecosystem forcing, and dissolved oxygen time series, due to its important role \red{in} biological processes, and 
also for the probable growing importance of this parameter to assess the quality of coastal
waters, in the framework of European directives \citep{Best2007}.

These physical and biogeochemical time series are nonlinear, \red{and} 
non-stationary, and may possess interactions at different scales.
In order to consider their multi-scale dynamic properties and explore their correlations at different scales, the Empirical Mode Decomposition (EMD) framework is applied here \citep{Huang1998EMD}. 

EMD and the associated Hilbert spectral analysis (resp. Hilbert-Huang Transform, HHT) have already been applied  in marine sciences, \citep{Hwang2003,Datig2004performance,Veltcheva2004,Schmitt2009JMS}. For example, \cite{Hwang2003} applied the HHT method to  ocean wave data. They found that the HHT method detects more energy in  lower frequencies, leading to  a lower average frequency  in HHT spectra than using  the Fourier framework. \cite{Datig2004performance} showed that the HHT method can used to study  nonlinear waves using instantaneous frequencies. \cite{Schmitt2009JMS} applied the HHT method to characterize the scale invariance of velocity fluctuations in the surf zone. They observed that the scale invariance holds for almost two decades of time scales.

In the following,  the methodology is presented and then its application is illustrated on the chosen time series. Section \ref{sec:method} presents the Hilbert-Huang Transform method and the \red{fairly} recent
time dependent intrinsic correlation; section \ref{sec:data} presents the data base;  section \ref{sec:results} presents the analysis of the intrinsic correlation and section \ref{sec:conclusion} draws the main conclusion of this paper.


\section{Hilbert-Huang Transform and Time Dependent Intrinsic Correlation}\label{sec:method}

In this section,  the Hilbert-Huang Transform  and the empirical mode 
decomposition based time dependent intrinsic correlation are presented.  These time series analysis techniques have been applied, since their introduction in 1998 \citep{Huang1998EMD}, in several thousand different studies in  natural and applied sciences. Here  the main idea is recalled \red{but} the method is not presented in too much detail.

The HHT consists \red{of} two steps. The first step is the so-called \lq{}empirical mode 
decomposition\rq{}, \red{ which} separates a multi-scale time series into a sum of intrinsic mode 
functions without \textit{a priori} basis assumption \citep{Huang1998EMD,Flandrin2004EMDa}. In the second step, the Hilbert spectral analysis is applied to each mode 
function to extract the time-frequency information.  The so-called Hilbert spectrum and the corresponding Hilbert marginal spectrum are then introduced to characterize the time-frequency distribution of a given time series \citep{Huang1998EMD,Huang2009PHD,Huang2008EPL,Chen2010AADA}.

\subsection{Empirical Mode Decomposition}

Empirical Mode Decomposition is  a fully adaptive technique  to   study
the nonlinear and non-stationary properties of time series \citep{Huang1998EMD,Huang1999EMD,Flandrin2004EMDa,Huang2011PRE}. 
The main idea of EMD is to locally separate  a given multi-scale signal into  a sum of a local trend and a local detail, respectively, for  a low frequency part and a high frequency part \citep{Rilling2003EMD}. 
The latter  is called  \red{the} Intrinsic Mode Function 
 (IMF), and the former is called the residual.  The procedure is repeated to the residual, considered as a new times series, extracting a 
 new IMF using a spline function, and obtaining a new residual until no more IMF can be extracted \citep{Huang1998EMD,Huang1999EMD,Rilling2003EMD,Flandrin2004EMDb}. 
 The EMD method then expresses a multi-scale time series as the sum of a finite number 
 of IMFs and a final residual  \citep{Huang1998EMD,Flandrin2004EMDb}. 

To be an IMF, an approximation to the so-called mono-component signal, it  must satisfy the following two conditions: (\romannumeral1)
 the difference between the number of local extrema and the number 
 of zero-crossings must be zero or one; (\romannumeral2) the running mean value 
 of the envelope defined by the local maxima and the envelope 
 defined by the local minima is zero \citep{Huang1998EMD,Huang1999EMD,Rilling2003EMD}. 
 The so-called Empirical Mode Decomposition algorithm is then proposed to  decompose a signal into IMFs \citep{Huang1998EMD,Huang1999EMD,Rilling2003EMD}:
 \begin{itemize}
 \item[1]   identify the local extrema of the signal $x(t)$;
 \item[2]   construct upper envelope  $e_{\max}(t)$ by using the  local maxima through a cubic spline interpolation (other interpolations
 are also possible).
Construct a lower envelope  $e_{\min}(t)$ by using the local minima;
 \item[3]   define  the mean value $m_1(t)=(e_{\max}(t)+e_{\min}(t))/2$;
 \item[4]   remove the mean value from the signal, providing the local detail
 $h_1(t)=x(t)-m_1(t)$;
 \item[5]   check if the component  $h_1(t)$  satisfies
 the above conditions to be an IMF.  If yes, take it as the first IMF 
 $C_1(t)=h_1(t)$. This IMF mode is then removed from the original signal and the first residual,
 $r_1(t)=x(t)-C_1(t)$  is taken as the new series in step 1.  If $h_1(t)$  is
 not an IMF, a procedure called \red{the} ``sifting process'' is applied as many times as
 necessary to obtain an IMF (not detailed here).
  \end{itemize}
  
 By construction, the number of extrema decreases when going from one
 residual to the next; the above algorithm ends when the residual has
 only one extrema, or is constant, and in this case no more IMF can be
 extracted; the complete decomposition is then achieved in a finite
 number of steps.
 The analyzed signal $x(t)$ is finally written as the sum of mode time series $C_i(t)$ and
 the residual $r_{n}(t)$:
\begin{equation}
x(t)=\sum_{i=1}^{N}{C_{i}(t)}+r_{n}(t)
\end{equation}
Based on a dyadic filter bank property of the EMD algorithm, the number of IMF modes is estimated as 
\begin{equation}
N\le \log_2(L)
\end{equation}
where $L$ is the length of the data in points \citep{Flandrin2004EMDb,Wu2004EMD,Huang2008EPL}.
Unlike Fourier based methodologies,  e.g., Fourier analysis, wavelet transform, etc., this method does not  define the basis \textit{a priori} \citep{Huang1998EMD,Huang1999EMD,Flandrin2004EMDa}. It thus possesses \red{full adaptability} and is very suitable for non-stationary and nonlinear time
series analysis \citep{Huang1998EMD,Huang1999EMD}. 

\subsection{Hilbert Spectral Analysis}

To characterize the time-frequency distribution of the IMF mode,  a complementary analysis technique namely Hilbert spectral analysis (HSA)  is then applied to each IMF mode to extract the local frequency information  \citep{Long1995,Huang1998EMD,Huang1999EMD,Huang2009PHD,Huang2011PRE}. In this complementary
step,  the Hilbert transform is used to construct the analytical signal, i.e.,
\begin{equation}
\tilde{C}(t)=C(t)+j\frac{1}{\pi}P\int_{-\infty}^{+\infty}\frac{C(t\rq{})}{t-t\rq{}}dt\rq{}\label{eq:analytical}
\end{equation}
in which $P$ is the Cauchy principle value \citep{Cohen1995,Long1995,Huang1998EMD,Flandrin1998}.
The above equation can be rewritten as 
\begin{equation}
\tilde{C}(t)=\mathcal{A}(t)\exp(j\theta(t))
\end{equation}
in which $\mathcal{A}(t)=\vert \tilde{C}(t) \vert$ is the modulus  and $\theta(t)=\arctan (\mathrm{IM} (\tilde{C}(t))/\mathrm{RE}( \tilde{C}(t)))$ is the instantaneous phase function \citep{Cohen1995,Long1995,Huang1998EMD,Flandrin1998,Huang2009PHD,Huang2011PRE}.
The instantaneous frequency is then defined as
\begin{equation}
\omega(t)=\frac{1}{2\pi}\frac{d \theta(t)}{dt}
\end{equation}
Note that the instantaneous frequency $\omega$ is very local since the Hilbert transform is a singularity transform and the differential operator is used to define the frequency $\omega$ \citep{Huang1998EMD,Huang1999EMD,Huang2009PHD}.  It \red{was} found experimentally that the Hilbert-based methodology is free with the Heisenberg-Gabor uncertainty and can be used to describe  nonlinear distortions by using an intrawave-frequency-modulation mechanism, in which the frequency can be varied with time in one period \citep{Huang1998EMD,Huang2005EMDa,Huang2011PRE}. Therefore, the method is free with high-order harmonic components, which are  required in Fourier-based methods to capture the non-stationary and nonlinear characteristics of the data \citep{Huang1998EMD,Huang1999EMD,Huang2005EMDa,Huang2011PRE}.

Note that several methods exist, that could be applied to estimate the instantaneous 
frequency, e.g., direct  quadrature, teager energy operator, etc.;  see more detail in 
\citet{Huang2009AADA}. In practice,  the Hilbert method already provides a good 
estimation of $\omega$ in a statistical sense \citep{Huang2010PRE,Huang2011PRE}.  

 A Hilbert spectrum $H(\omega,t)=\mathcal{A}^2(\omega,t)$ can be designed to represent the energy of the
original signal
as \red{a} function of   frequency $\omega$ and time. It can be
taken as the best local fit to $x(t)$ using an amplitude and phase varying
trigonometric function \citep{Huang2005EMDa}. This corresponds to a high resolution fit
both in  physical space and frequency space
\citep{Long1995,Huang1998EMD}.  
 Then, the Hilbert marginal
$h(\omega)$ can be defined as
\begin{equation}
h(\omega)=\frac{1}{T}\int_{T} H(\omega,t)\,dt\label{eq:hms}
\end{equation}
in which $T$ is the time period to calculate the spectrum.
This is comparable to the power spectrum in
Fourier analysis, but it has a quite different definition and the resulting
curve \red{would be} expected to be different from the Fourier $E(f)$ curve  if the times series
is nonlinear and non-stationary \citep{Huang1998EMD}. In fact, here the definition of
instantaneous frequency is different \red{from} the one in \red{the} Fourier frame
and the interpretation and detailed physical meaning of $h(\omega)$ is
still to be fully characterized.

Another way to estimate $h(\omega)$ is based on the joint probability density function (pdf) $p(\mathcal{A},\omega)$ of amplitude $\mathcal{A}$ and the instantaneous frequency $\omega$ from all IMF modes \citep{Huang2008EPL,Huang2009PHD,Huang2010PRE,Huang2011PRE}. It is written as
\begin{equation}
h(\omega)=\int_{-\infty}^{+\infty}p(\mathcal{A},\omega)\mathcal{A}^2 d \mathcal{A}\label{eq:hms2}
\end{equation}
The combination of the EMD and HSA is then called Hilbert-Huang Transform \citep{Huang2005EMDa}. It has been applied successfully in various research fields to characterize the energy-time-frequency in different topics \citep{Zhu1997,Echeverria2001,
Coughlin2004,Loutridis2004,Janosi2005,Molla2006a,Molla2006,Huang2008EPL,Huang2009Hydrol,Schmitt2009JMS,Chen2010AADA}, to \red{cite} a few.

\subsection{Time Dependent Intrinsic Correlation}
The classical global expression for the correlation,
applied to non-stationary time series, may 
\red{distort} the true  cross correlation information between  time series and provide 
unphysical interpretation \citep{Hoover2003,Rodo2006,Chen2010AADA}. An 
alternative way, consistent with the possible non-stationarity of the time series, is to estimate the cross correlation coefficient by using a sliding 
window or a scale dependent correlation technique 
\citep{Papadimitriou2006,Rodo2006}. But the main problem of these techniques is 
to determine how large the window should be  \citep{Chen2010AADA}. 
Time series from natural sciences possess fluctuations in a whole range of scales, and the data
characterization must take into account this scale question  \citep{Cohen1995,Huang1998EMD,Flandrin1998}.   The estimation of the cross correlation
between time series, in a multi-scale framework, may use a window based on the
local characteristic scale given by the data itself.  

Recently, \citet{Chen2010AADA} proposed  
to use the EMD to estimate an adaptive window in order to calculate a so-called time-dependent 
intrinsic correlation (TDIC). The TDIC of each pair of IMFs is defined as follows:
\begin{equation}
R_{ij}(t_k^n\vert t_w^n)=\mathrm{Corr}\left(C_{1,i}(t^n_{w})C_{2,j}(t^n_{w}) \right) \textrm{ at any }t_k
\end{equation}
where Corr is the cross correlation coefficient of two time series and $t^n_{w}=[t_k -n 
t_d/2 : t_k +n t_d/2]$ is the sliding window. The minimum sliding window size for the local correlation estimation  is 
chosen as $t_d = \max(T_{1i}(t_k),T_{2j}(t_k))$, where $T_{1i}$ and $T_{2j}$ are the 
instantaneous periods $T=\omega^{-1}$, and $n$ is any positive real number 
\citep{Chen2010AADA}.   The efficiency of this approach to characterize the relation between two time series for several problems has been shown in \citet{Chen2010AADA}. Here we use this new method (with $n\ge1$) to estimate cross correlations between
coastal marine time series, where the latter can be considered as typical nonlinear and
non-stationary data. 


\section{Presentation of the experimental database}\label{sec:data}

\begin{figure}  [!htb]
\centering
  \includegraphics[width=0.85\textwidth,clip]{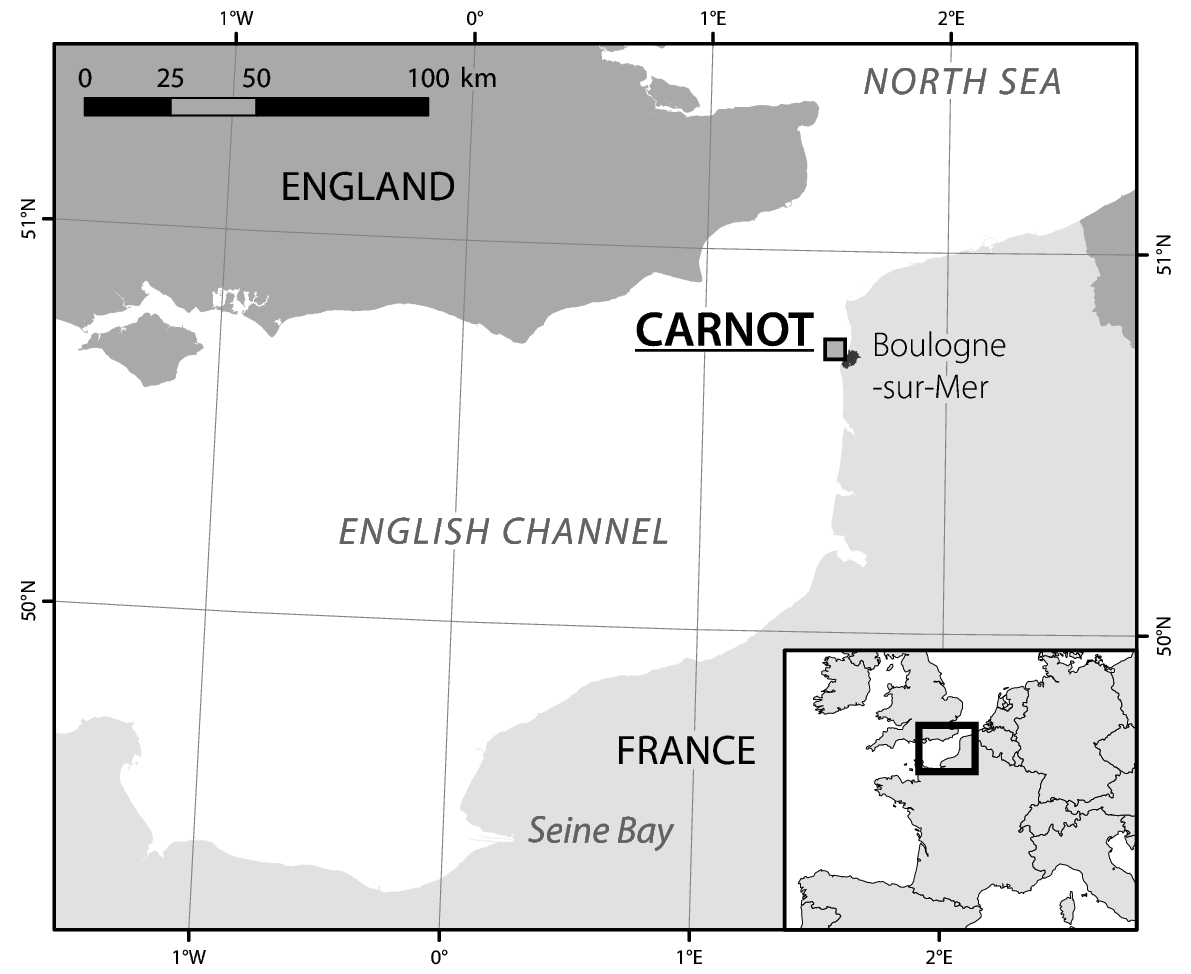}
\caption{A map showing the measurement location Boulogne-sur-mer (France)  at position 50.7404 N, 1.5676 W, in the eastern English Channel.}\label{fig:map} 
\end{figure}

The time series analyzed here belong to the MAREL network
 (Automatic monitoring network for littoral environment, Ifremer, France\footnote{http://www.ifremer.fr/marel}). Such \red{a} system is based on the deployment of moored buoys equipped with physico-chemical measuring devices, working in continuous and autonomous conditions \citep{Woerther1998,Blain2004}. These stations use automatic systems for seawater analysis and real time data transmission. They record \red{several} parameters, such as temperature, salinity, dissolved oxygen, pH, \red{and} turbidity, with a fixed time resolution. The measuring station used here, the MAREL Carnot station, is situated in the eastern English Channel in the coastal waters of
Boulogne-sur-mer (France)  at position 50.7404 N, 1.5676 W (Figure\,\ref{fig:map}) and records data with a 20 minutes resolution \citep{Zongo2011b}. 
Water depth at this position varies between 5 and 11 m; the measurements are done using a floating system inserted in a tube, 1.5m below the surface.
Statistical and scaling properties of MAREL automatic measurements in the English Channel have been studied previously, considering temperature time series  \citep{Dur2007} and  turbidity, oxygen and pH \citep{Schmitt2008} in the Seine estuary, and pH fluctuations in both the Seine estuary and Boulogne-sur-mer waters \citep{Zongo2011a}.

Among other parameters, the measuring buoy provides temperature, salinity and dissolved oxygen (DO)  concentration data, as well as percentage
of DO relative to the dissolved oxygen at equilibrium ($SaO_2$) for the same temperature and salinity.
The DO time series is directly linked to temperature due to the solubility equation given below, whereas $SaO_2$ is free from such direct influence. In the following,  the latter dissolved oxygen series is considered.
The oxygen saturation percentage equation is given by
Garcia and Gordon, as a nonlinear function
of the form \citep{Garcia1992}:
\begin{equation}
 SaO_2 = \frac{100DO}{\exp\left(P_1(T)+SP_2(T)+C_0 S^2 \right)}
 \label{eqxx}
\end{equation}
where  the denominator is the nonlinear fit expressing
the oxygen solubility, $S$ and $T$ are salinity and temperature and $P_1$ and
$P_2$ two polynomial developments ($P_1$ of degree five, and $P_2$ of degree three).
Their coefficients are given in  \citet{Garcia1992}. $SaO_2$ data have thus no dimension and values exceeding $100 \, \%$ correspond to supersaturation  while
values below $100 \, \%$ correspond to undersaturation associated to oxygen depletion; values lower than $50 \, \%$ are found for anoxic waters which can be dangerous for many  species.

\begin{figure}  [!htb]
\centering
  \includegraphics[width=0.7\textwidth,clip]{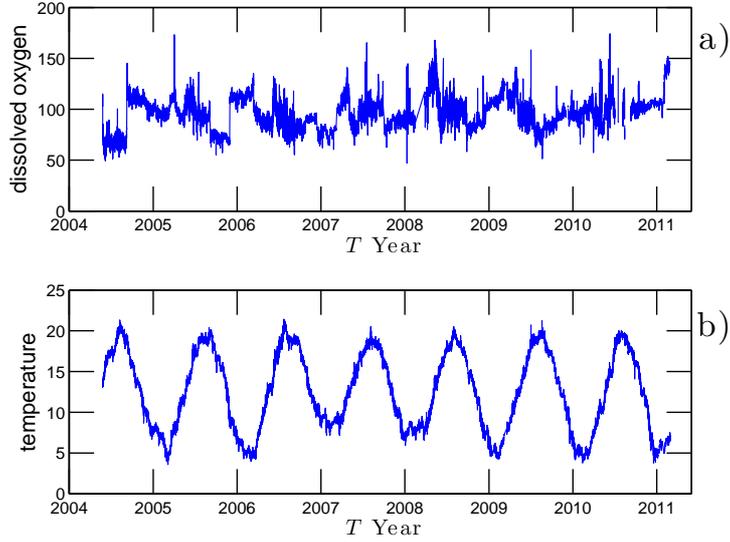}
\caption{The observed dissolved oxygen a) and the temperature b) on the time period 25th May 2004 to 1st March 2011. Note that the sampling interval is not always 20 minutes, due to missing data. The mean period is 24 minutes for both data sets. A moderate warming winter is observed  in  2007 and  2008. The observed temperature during the winter is  $2\,^{\circ}\mathrm{C}$ higher than usual. Note that  a strong riding-wave phenomenon  is observed from January to April 2008. The observed riding-wave will cause the mode mixing problem in the EMD decomposition. }\label{fig:data} 
\end{figure}

\begin{figure}  [!htb]
\centering
  \includegraphics[width=0.7\textwidth,clip]{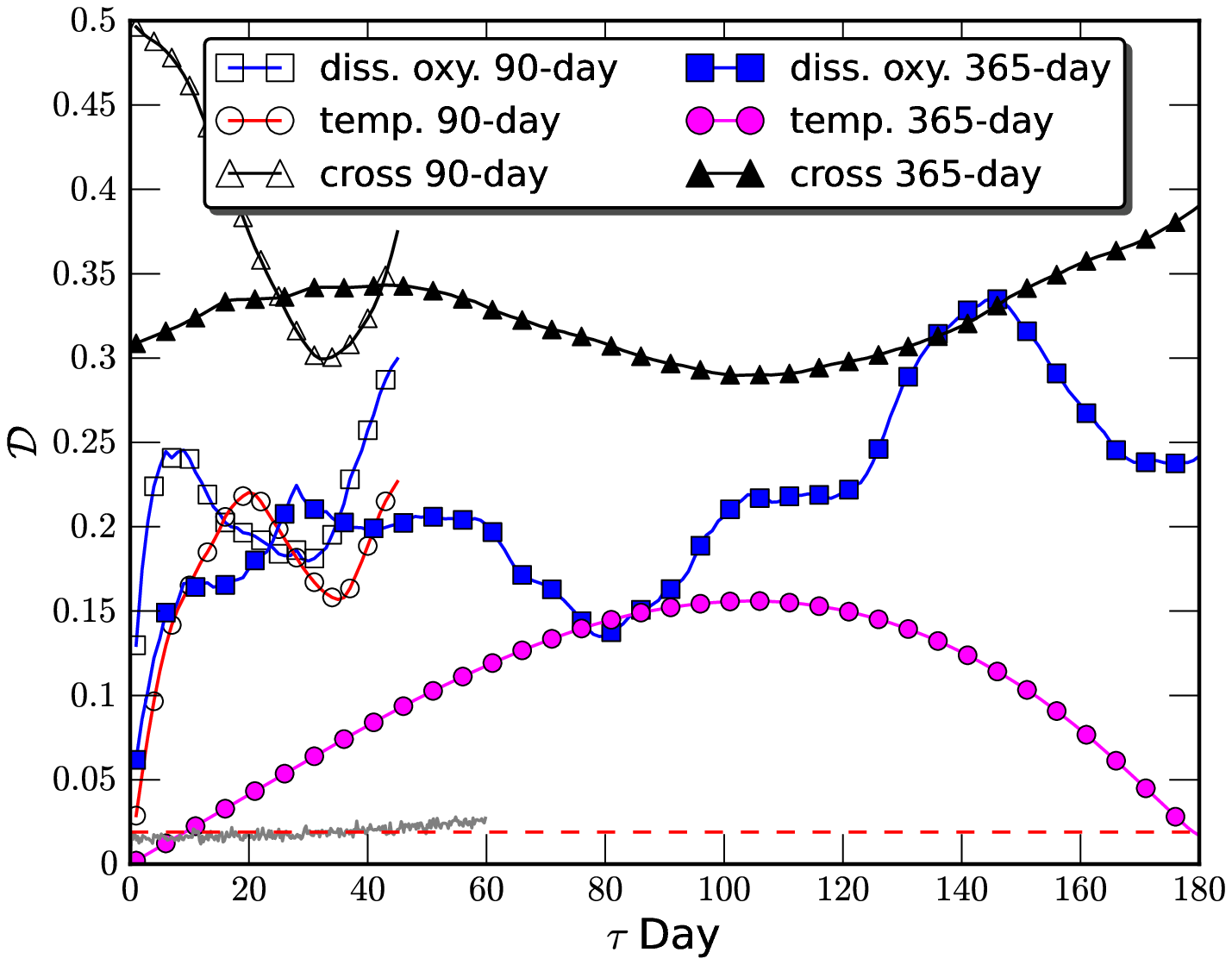}
\caption{ Statistical test of the degree of stationary $\mathcal{D}$ for the autocorrelation and cross-correlation of two data sets with a sliding window $t_w=90$ days and $t_w=365$ days. The test for white noise is illustrated as a horizontal dashed line, showing the validation of the definition of the degree of stationary. Note that the measured $\mathcal{D}$ is strongly dependent on the windows size $t_w$. The curves indicate that none of the series are stationary but that the temperate data set is more stationary than the dissolved oxygen one. }\label{fig:DOS}
\end{figure}

Due to occasional failures in the measuring system, and to regular maintenance, the time series sampling frequency is not always 20 minutes; there are \red{frequently}
 missing data and the 
 mean sampling period is 24 minutes for both data sets. 
A total of 145,587 data points for dissolved oxygen collected from 25th May 2004 
to 1st March 2011 and 155,825 data points for sea temperature collected from 24th March
2004 to 1st March 2011 have been analyzed.  Figure \ref{fig:data} shows the two data sets.  
The temperature data displays a strong annual cycle.  These two 
data sets have been analyzed on the same time period 25th May 2004 to 1st March 2011.  There is a warm winter in 2007 and 2008, showing also a weak oscillation of the temperature of the sea water. 

The degree of stationary  which was introduced by \citet{Chen2010AADA} is first calculated, i.e.,
\begin{equation}
\mathcal{D}(\tau)=\mathrm{std}\left(R(t^n_k,\tau )\right)\label{eq:dos}
\end{equation}
in which $R(t^n_k,\tau)$ is an autocorrelation function with a sliding window $t^n_k$ and time delay $\tau$. The value of $\mathcal{D}$  depends on the window size and the time delay $\tau$. It varies from $0$ to $1$ and 
\red{characterizes} the deviation from a stationary process \citep{Chen2010AADA}: 0 for exactly stationary processes and the larger the value, the more it is 
non-stationary.
Figure \ref{fig:DOS}  shows the measured degree of stationary $\mathcal{D}$ 
for the autocorrelation and cross-correlation with a sliding window size of $t_w=90$ days and $t_w=365$ days. 
To show the validation of  Eq.\eqref{eq:dos}, a test for white noise is 
performed, which is illustrated by a dashed line. As expected a small value of  
$\mathcal{D}$ is found for white noise. These results applied on two 
experimental time series indicate that both are non-stationary and that the 
temperature is more stationary than the dissolved oxygen.

\section{Analysis Results}\label{sec:results}

\subsection{Empirical Mode Decomposition Result}

\begin{figure}  [!htb]
\centering
  \includegraphics[width=0.7\textwidth,clip]{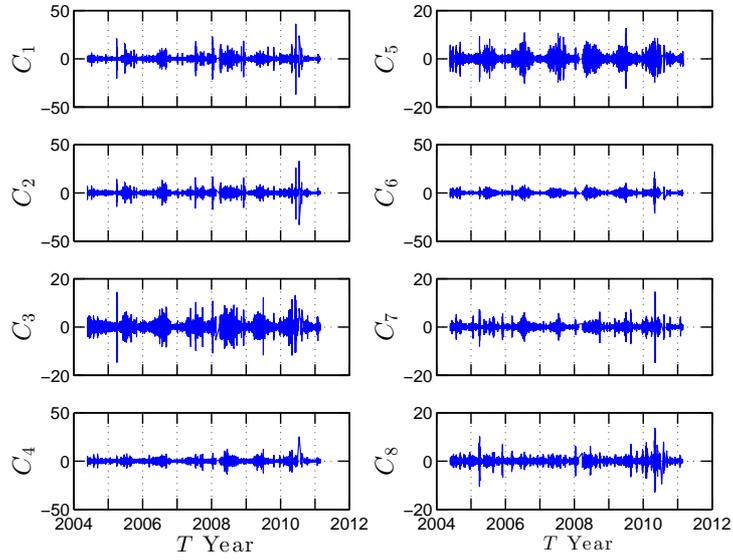}
\caption{ The first-eighth IMF modes for dissolved oxygen. The time scale is increasing with the mode index $n$.}\label{fig:imfDOa}
\end{figure}

\begin{figure}  [!htb]
\centering
  \includegraphics[width=0.7\textwidth,clip]{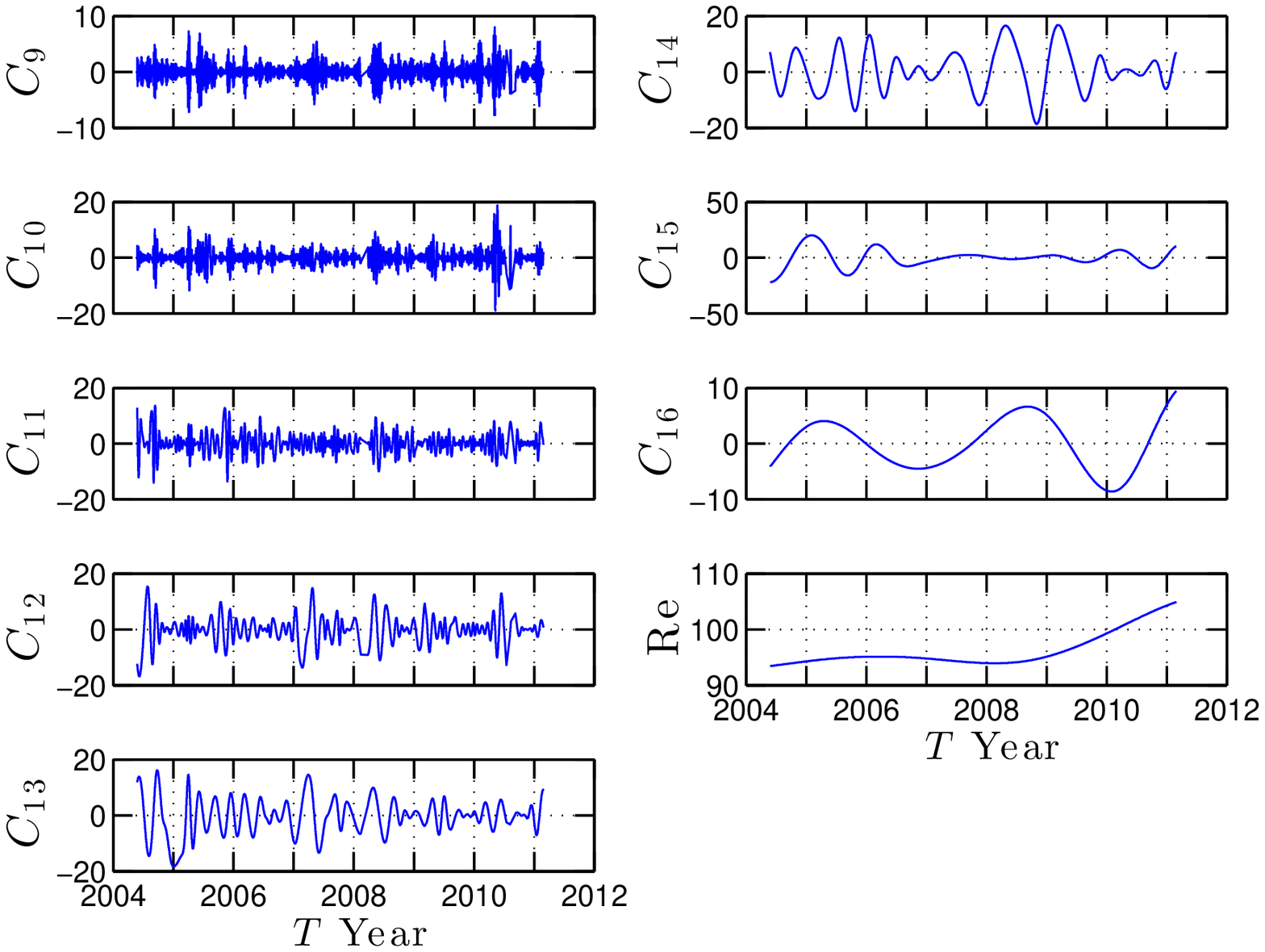}
\caption{ The last-eighth IMF modes for dissolved oxygen. The time scale is increasing with the mode index $n$.}\label{fig:imfDOb}
\end{figure}

\begin{figure}  [!htb]
\centering
  \includegraphics[width=0.7\textwidth,clip]{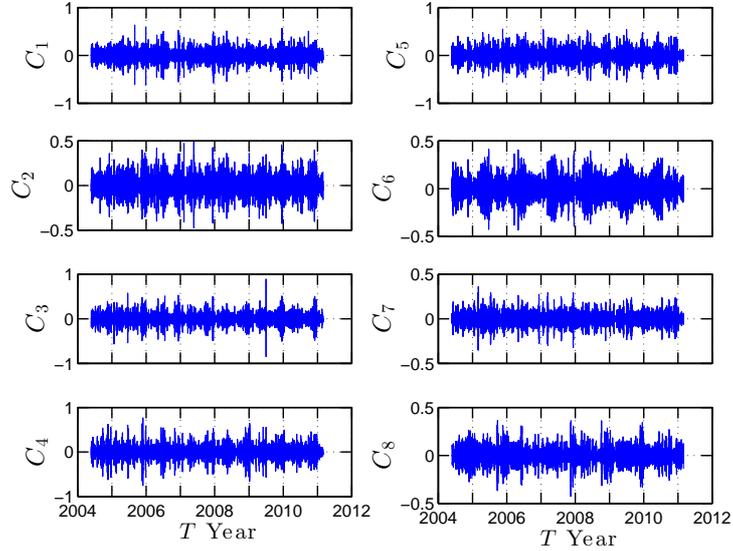}
\caption{ The first-eighth IMF modes for temperature. The time scale is increasing with the mode index $n$.}\label{fig:imfTa}
\end{figure}

\begin{figure} [!htb]
 \centering
  \includegraphics[width=0.7\textwidth,clip]{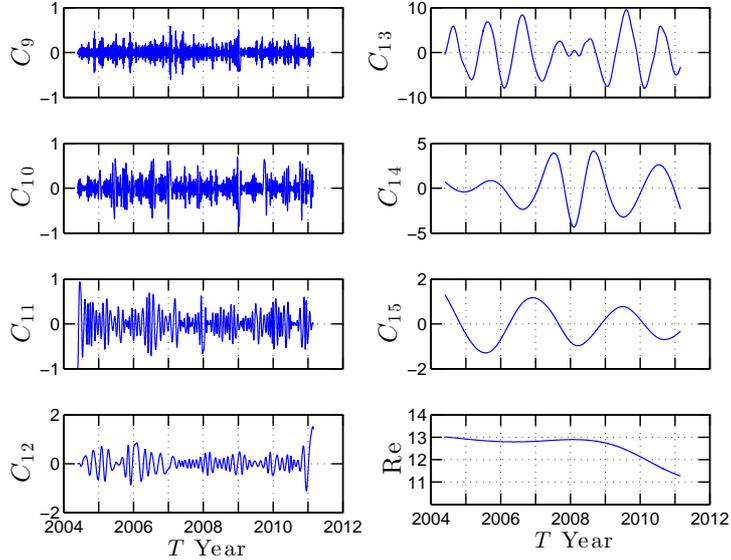}
\caption{ The last-seventh IMF modes together with the trend term for temperature. The time scale is increasing with the mode index $n$. Note that there is a serious mode mixing problem for the $13${th} IMF mode around $2008$. The mean period of this mode is around 1 year.  The mode mixing is caused by a  riding-wave phenomenon from January 2008 to April with a typical time scale of 3 months, see Fig.\,\ref{fig:data}\,b. }\label{fig:imfTb}
\end{figure}

\begin{figure} [!htb]
 \centering
  \includegraphics[width=0.7\textwidth,clip]{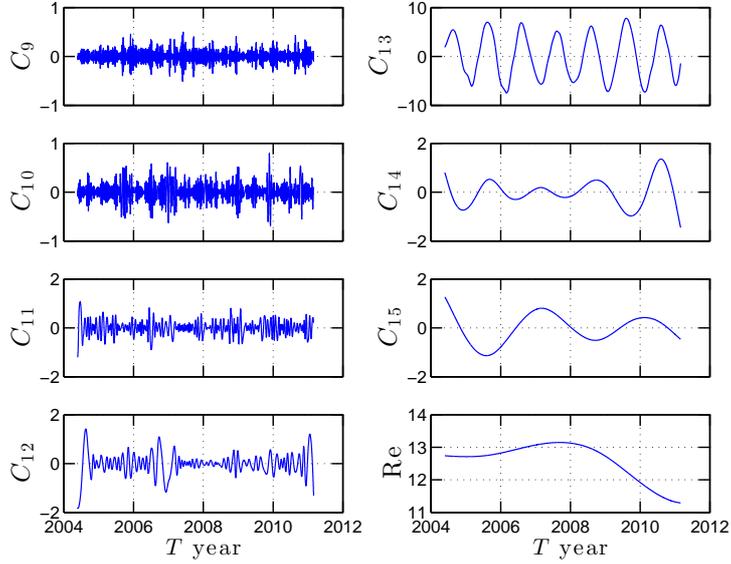}
\caption{ The last-seventh IMF modes together with the trend term for temperature by removing the riding-wave. A better annual cycle, i.e., $C_{13}$, is obtained.  Note that the trend is not influenced by removing the riding-wave. }  \label{fig:imfTc}
\end{figure}

\begin{figure}  [!htb]
\centering
  \includegraphics[width=0.7\textwidth,clip]{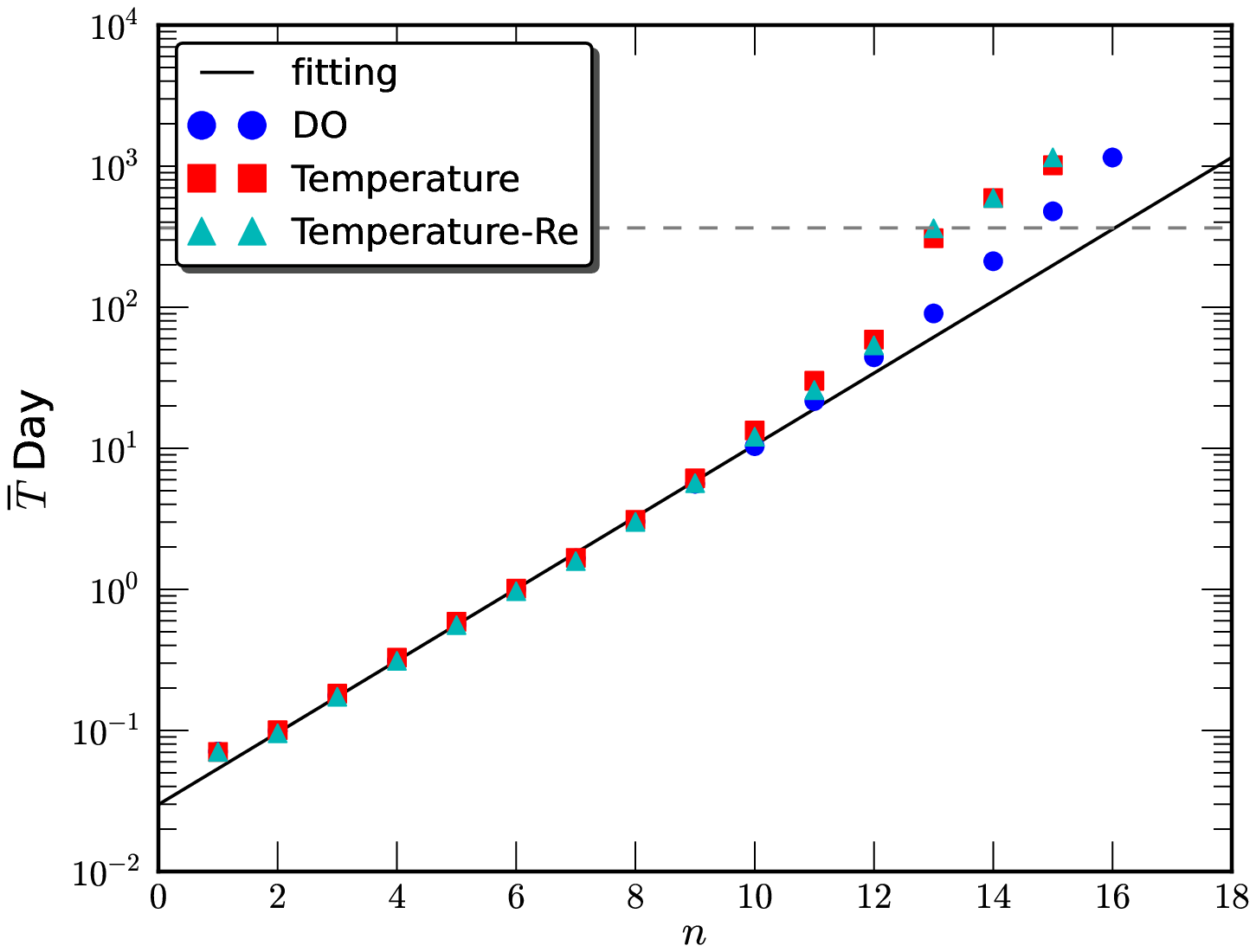}
\caption{ Mean period of  IMF modes obtained for dissolved oxygen ($\ocircle$), temperature with ($\square$) and without ($\triangle$) riding-wave, respectively. The annual cycle is indicated by a dashed line. A better annual cycle for the temperature is retrieved by removing the riding-wave. A exponential law is observed for both data sets with an exponent $1.78\pm0.04$, indicating a quasi dyadic-like filter bank of the EMD algorithm. Note that the first twelve IMF modes for both data sets approximately have the same mean period.}\label{fig:mp}
\end{figure}

The EMD algorithm is applied to the two observed data sets \red{for} 
the same time period without interpolating the data into a 
regular time interval, since in the EMD step the information is contained in the local extrema points and can be applied to \red{the} time series with irregular time intervals \citep{Huang1998EMD,Flandrin2004EMDa}. After the EMD decomposition, there are respectively, 16 and 15 IMF modes with one residual for the dissolved oxygen and temperature.
Figures \ref{fig:imfDOa} to \ref{fig:imfTb} show the 
corresponding IMF modes. Note that the time scale is 
increasing with the index  $n$: the number of the IMF mode. 
 The IMF modes capture the  local variation of the time scales. 
 Note that  there is a serious mode mixing problem for the $13$th IMF mode of the temperature around 2008, see Fig.\,\ref{fig:imfTb}. This is caused by a 
 riding-wave phenomenon  in the original temperature data \citep{Huang1998EMD}, see Fig.\,\ref{fig:data}\,b.  To eliminate the mode mixing 
 problem, a noise assistant method, namely Ensemble Empirical Mode 
 Decomposition (EEMD) could be applied to the data \citep{Wu2009}. However, 
 practically speaking,  the EEMD will introduce additional scales  and  bring bias  
 \red{to} the estimation of the Hilbert spectrum $h(\omega)$. Therefore  the 
 riding-wave was removed manually. Figure \ref{fig:imfTc} shows the last-seventh IMF modes without the riding-wave: it provides a better $13$th IMF mode. Note that the trend term is not influenced by the riding-wave.
The mean period of the each IMF mode is estimated by calculating the local extrema points and zero-crossing points \citep{Rilling2003EMD,Huang2009Hydrol},  i.e.,
 \begin{equation}
 \overline{T}(n)=\frac{L}{N_{n,\max}+N_{n,\min}+N_{n,0}}\times 4
 \end{equation}
 in which $L$ is the length of the data and $n$ is the index.
 Figure \ref{fig:mp} shows the measured $\overline{T}$ in a 
 semilog plot. The first 12 IMF modes of the two data 
 sets approximately have the same mean period $\overline{T}$ and follow an 
 exponential law, i.e.,
 \begin{equation}
 \overline{T}(n)= \alpha\times \gamma^{n}
 \end{equation}
 in which $\alpha\simeq 0.035$ and $\gamma\simeq 1.78$ 
 obtained by using a least square 
 fitting algorithm.  This value is close to  $2$, which 
 indicates a quasi dyadic filter bank property of the EMD algorithm for these series, as found in
 other situations
 \citep{Wu2004EMD,Flandrin2004EMDb,Huang2008EPL,Huang2010PRE}. In other words, the mean scale
 of each mode series is 1.78 times the mean scale of the previous one.

\subsection{Hilbert Spectrum}

\begin{figure} [!htb]
\centering
  \includegraphics[width=0.7\textwidth,clip]{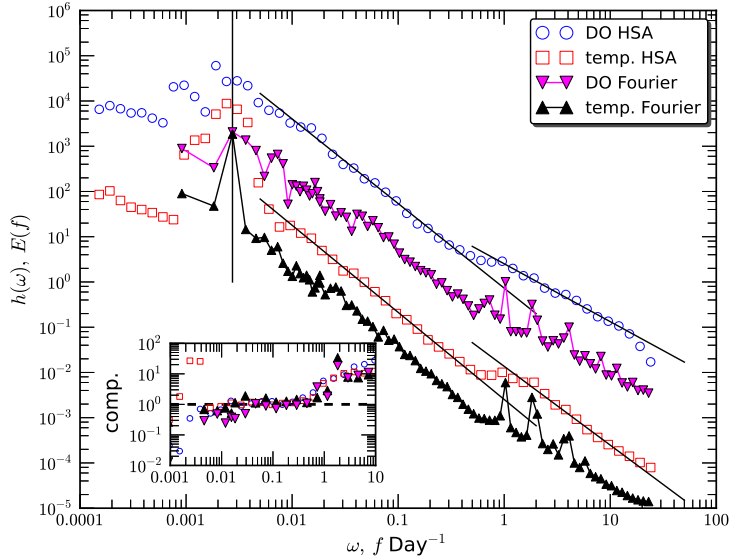}
\caption{ The Hilbert spectrum of dissolved oxygen ($\ocircle$) and temperature ($\square$),  in which the vertical line indicates the annual cycle. Power law behavior is observed for both curves on the range $0.01<\omega<0.5$ day$^{-1}$ and $2<\omega<20$ day$^{-1}$, corresponding to a time scale $2<T<100$ days and $1.2<T<12$ hours, respectively. The corresponding scaling exponents are $1.93\pm0.05$ and $1.87\pm0.08$ for large scales, and $1.68\pm0.10$ and $1.35\pm0.10$ for small scales respectively. For comparison, the Fourier power spectrum \red{is} also shown for both dissolved oxygen ($\triangledown$) and temperature ($\triangle$). The inset shows the compensated spectrum $h(\omega)\omega^2C^{-1}$ ($E(f)f^2C^{-1}$) with fitted $C$ to emphasize the observed power laws.} \label{fig:hms}
\end{figure}

With the IMF modes obtained from EMD algorithm,  the HSA  can be applied to each IMF mode. 
In this step, we first interpolate the IMF mode by using a Piecewise Cubic Hermite Interpolating 
Polynomial  (phcip) algorithm into a regular time interval with $dt=30$ minutes, which is slightly 
larger than the mean time interval of $24$ minutes. Then the HSA is applied to extract the joint pdf 
$p(\mathcal{A},\omega)$. The Hilbert marginal spectrum is then calculated by using Eq.\eqref{eq:hms2}. Figure \ref{fig:hms} shows the Hilbert marginal spectra for dissolved oxygen ($\ocircle$) and temperature ($\square$), in which the vertical line indicates the annual cycle.  A strong annual cycle is observed for the temperature data, which is consistent with the observation of the original temperature data, see Fig. \ref{fig:data} b.  It is emphasized here that for the temperature data with and without riding-wave around 2008, the Hilbert spectrum $h(\omega)$ are the same, showing the robustness of the present method.
Power law behavior is observed on the frequency range $0.01<\omega<0.5$ day$^{-1}$, corresponding to time scales $2<T<100$ days (resp. $T=\omega^{-1}$), and $2<\omega<20$ day$^{-1}$, corresponding to time scales $1.2<T<12$ 
hours. The measured scaling exponents are respectively $1.93\pm0.05$ and 
$1.87\pm0.08$ for the first scaling range of temperature and dissolved oxygen,   \red{and}
$1.68\pm0.10$ and $1.35\pm0.10$ for the second scaling range. The scaling 
exponents for the first scaling range are statistically close to each other. For the 
second scaling exponent, the temperature one is close to the Kolmogorov value 
$5/3$ \citep{Frisch1995}, implying that the variation of the temperature within one 
day is like a passive scalar.  For comparison, the corresponding Fourier power 
spectrum $E(f)$ is also presented. Both Hilbert and Fourier spectra show  strong 
annual  and daily cycles. However, there is no half-day cycle  observed  in the 
Hilbert spectrum. It might indicates that the half-day cycle is a harmonic from the 
nonlinear distortion of the daily cycle.  Power law behavior is also observed for  \red{the}
Fourier power spectrum on the same range of time scales $2<T<100$ days as for the 
Hilbert framework.  The corresponding scaling slope is close to $2$. To emphasize the 
observed power law behavior, the compensated spectra 
$h(\omega)\omega^2C^{-1}$ (resp. $E(f)f^2C^{-1}$) are shown in the inset. The observed plateau confirms the power law. Note that the observed $-2$ power law indicates that there might be a cascade process in the marine coastal environment \citep{Thorpe2005turbulent}. This should be investigated further in future studies. 
Let us  note that a shorter portion of this data set has
been studied in \citet{Zongo2011a}: for the years 2006 to 2008, with about 50,000 data points, a Fourier analysis has found a power-law slope of $1.66$ for scales between 1 year and 2 hours, where the fit was done for this whole range; it \red{should} be noted that different fits on separated portions of the frequencies give values closer to the Hilbert-based estimates. 

\subsection{Time Dependent Intrinsic Correlation}

\begin{figure} [!htb]
\centering
  \includegraphics[width=0.7\textwidth,clip]{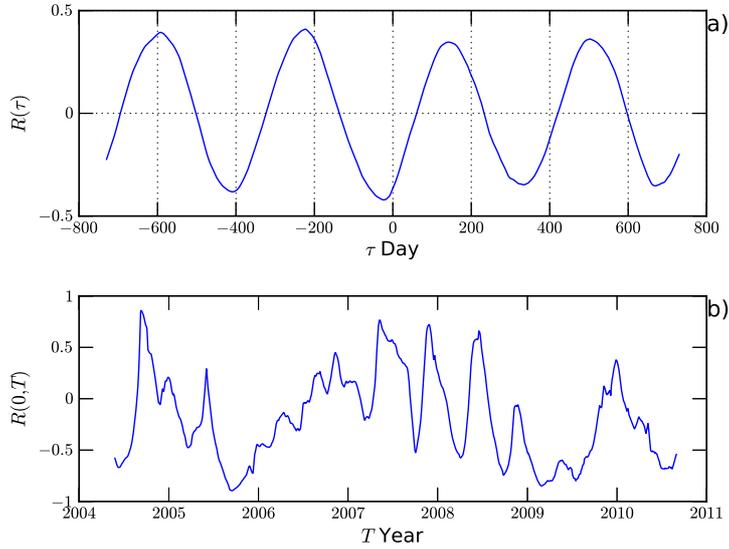}
\caption{ a) The global cross correlation coefficient $R(\tau)$ between the dissolved oxygen and the temperature. Note that there is a strong annual cycle with a phase difference  \red{of} 30 days.  The global correlation coefficient is $R(0)=-0.37\pm0.10$. b) The measured  cross correlation coefficient $R(0,t_w)$ with a sliding window $t_w=180$ days. It shows that the relation between dissolved oxygen and temperature \red{varies} from time to time, indicating a multi-scale property of such data sets. }\label{fig:globalcorrelation}
\end{figure}

\begin{figure} [!htb]
\centering
  \includegraphics[width=0.7\textwidth,clip]{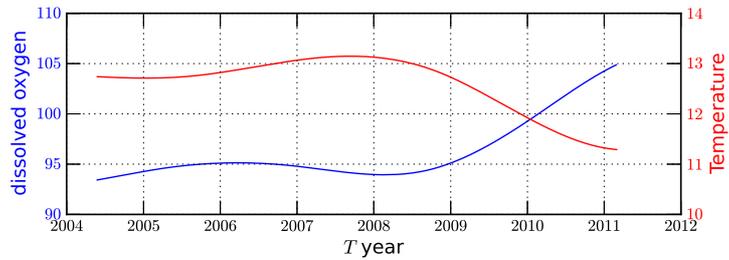}
\caption{ The trend from EMD for dissolved oxygen and temperature. The direct measurement of the cross correlation is $-0.96$, showing a global out-of-phase relation between the temperature and dissolved oxygen variation.  }\label{fig:trend}
\end{figure}

\begin{figure} [!htb]
\centering
  \includegraphics[width=0.85\textwidth,clip]{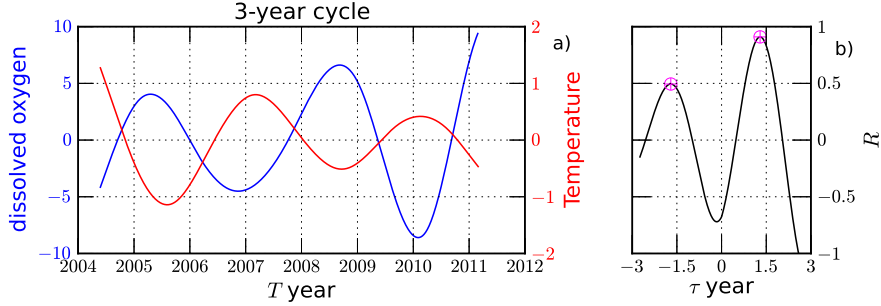}
\caption{ a) The IMF modes with a 3-year mean period. b) The measured cross correlation $R(\tau)$. They are negatively correlated with a phase difference  \red{of} around  2 months. The overall correlation coefficient is $R(0)=-0.68$. The mean period is determined by the distance between the two local extrema points ($\oplus$). }\label{fig:3year}
\end{figure}

\begin{figure} [!htb]
\centering
  \includegraphics[width=0.85\textwidth,clip]{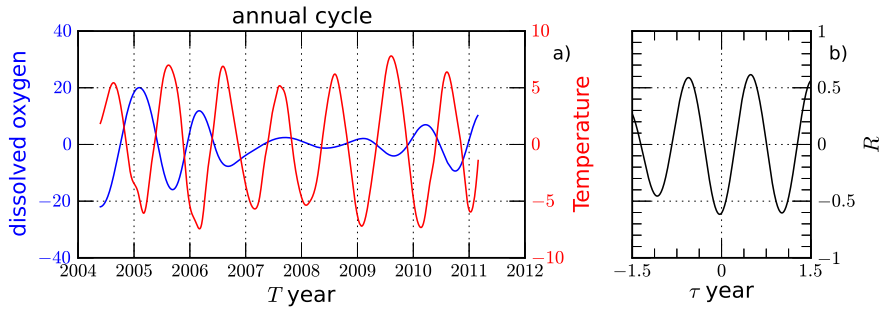}
\caption{ a) The annual cycle from EMD.  b) The measured cross correlation $R(\tau)$. The overall correlation coefficient is $R(0)=-0.60$ with  14 days phase difference.  Note that there is a weak period between 2007 and 2009 for both temperature and dissolved oxygen.   }\label{fig:annual}
\end{figure}

\begin{figure} [!htb]
\centering
  \includegraphics[width=0.7\textwidth,clip]{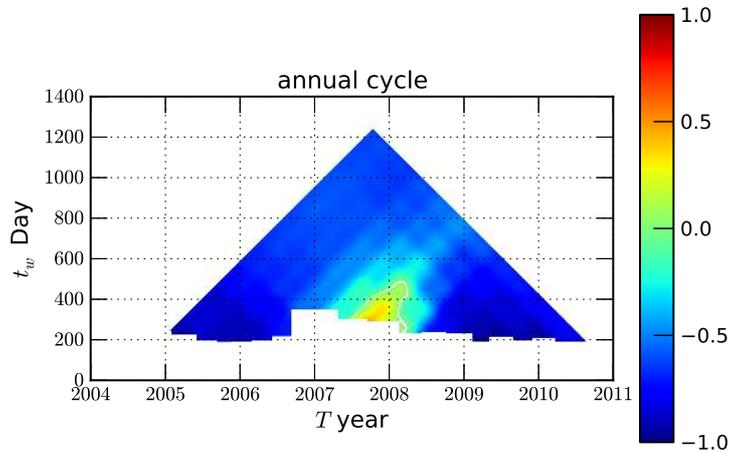}
\caption{ The measured TDIC for the annual cycle.   
It shows a transition from negative positive correlation between 2007 
and 2008. The overall correlation coefficient is -0.60, see Fig.\,\ref{fig:annual}\,b.   The hole is the $R$ \red{cannot} pass the $t-$test.  The horizontal axis is the location of the \red{centre} of the sliding window.}\label{fig:annualTDIC}
\end{figure}

\begin{figure} [!htb]
\centering
  \includegraphics[width=0.7\textwidth,clip]{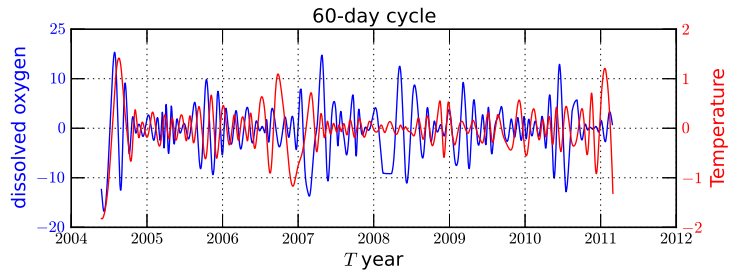}
\caption{ The 60-day cycle from EMD.  The overall correlation coefficient is -0.02 with  15 days phase difference.     }\label{fig:60day}
\end{figure}

\begin{figure} [!htb]
\centering
  \includegraphics[width=0.7\textwidth,clip]{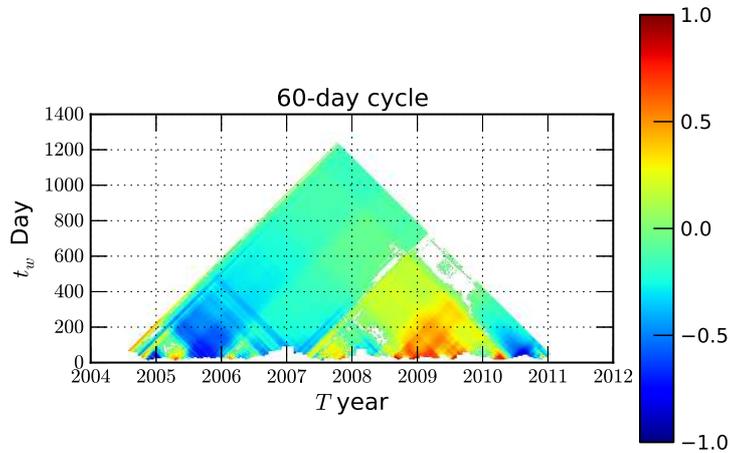}
\caption{ The measured TDIC for 60-day mean period.   There is a more than half year strong correlation pattern  around 2009 and 2010. The overall correlation coefficient is -0.02.   }\label{fig:60dayTDIC}
\end{figure}

\begin{figure} [!htb]
\centering
  \includegraphics[width=0.7\textwidth,clip]{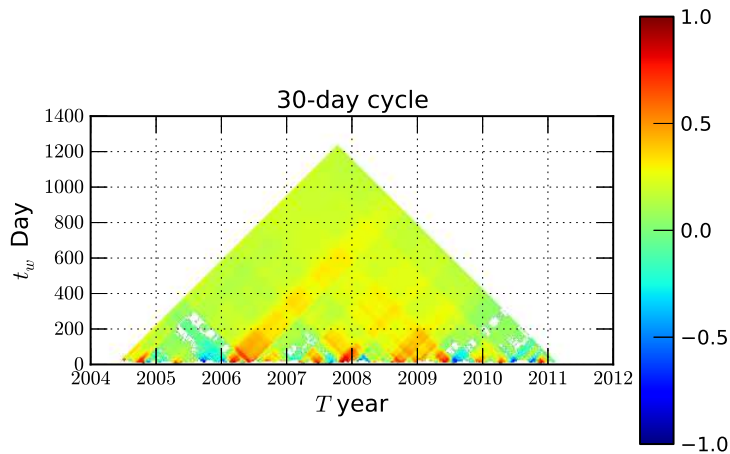}
\caption{ The measured TDIC for 30-day mean period on the range May 2004 $\sim$ January 2008.   There are more than half \red{a} years strong correlation patterns  between January 2006 and  June 2006. The overall correlation coefficient is 0.17.   }\label{fig:30dayTDIC}
\end{figure}

\begin{figure} [!htb]
\centering
  \includegraphics[width=0.7\textwidth,clip]{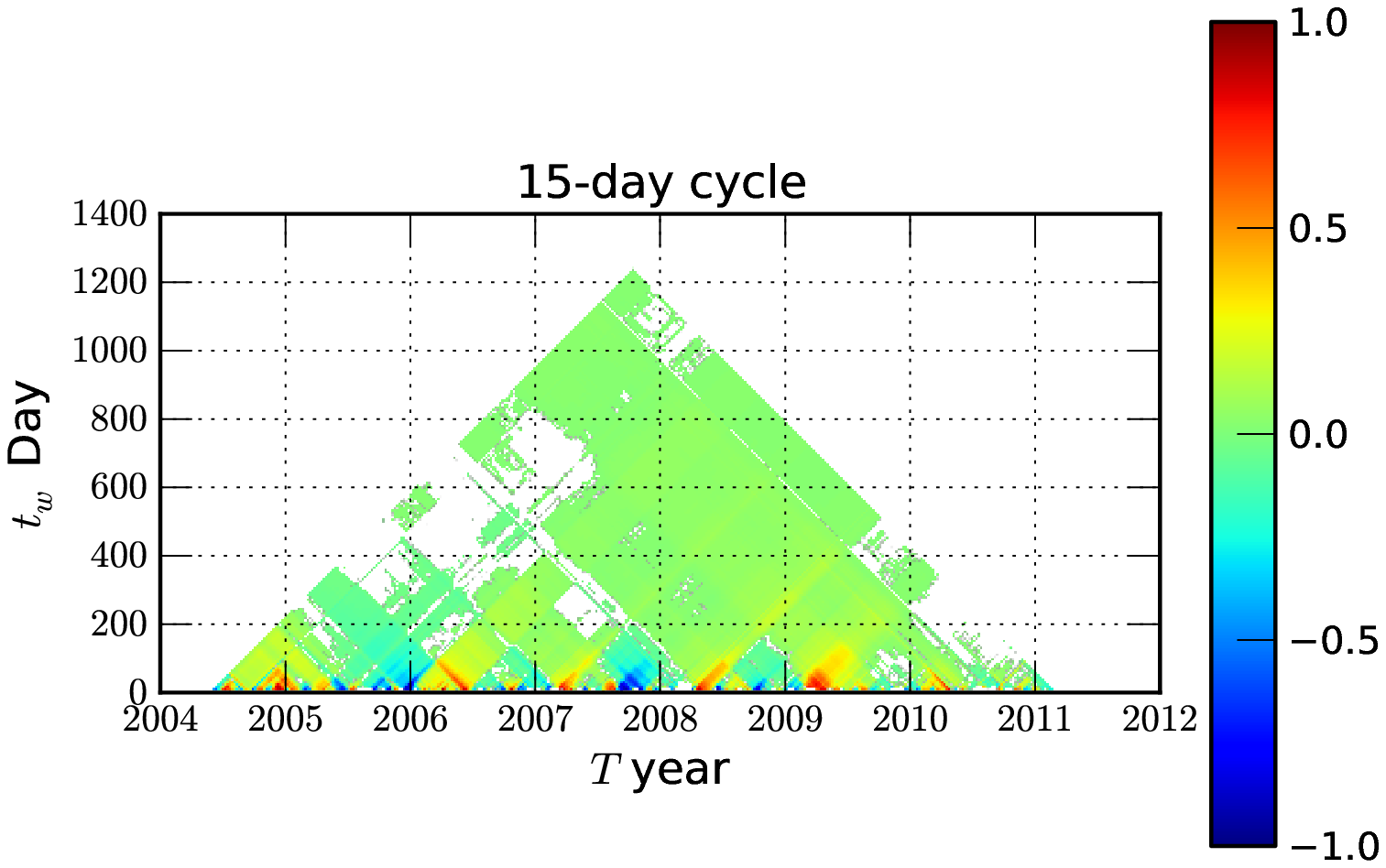}
\caption{ The measured TDIC for the 15-day mean period. The overall correlation coefficient is 0.10.   }\label{fig:15dayTDIC}
\end{figure}

\begin{figure} [!htb]
\centering
  \includegraphics[width=0.7\textwidth,clip]{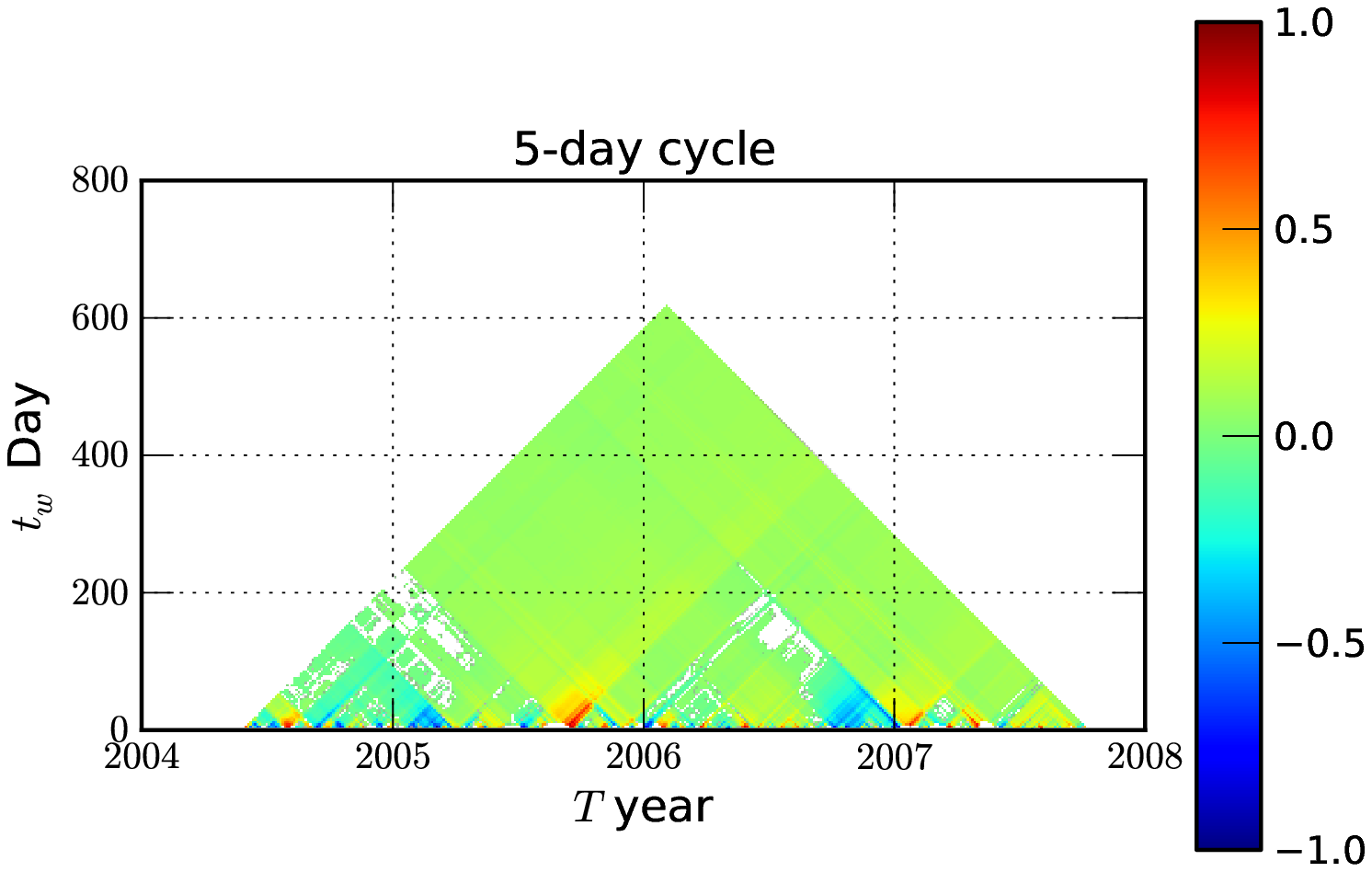}
\caption{ The measured TDIC for the 5-day mean period. The overall correlation coefficient is 0.1.   }\label{fig:5dayTDIC}
\end{figure}

\begin{figure} [!htb]
\centering
  \includegraphics[width=0.7\textwidth,clip]{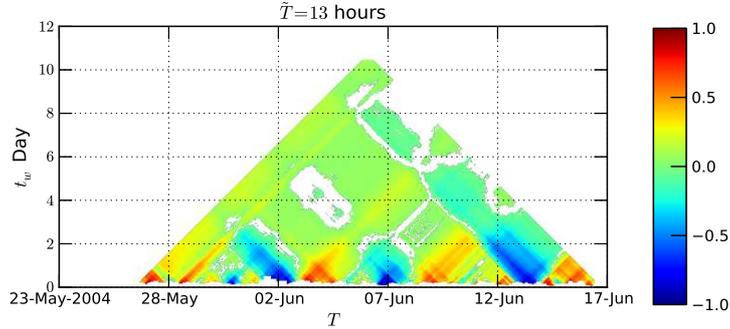}
\caption{ The measured TDIC for the 13-day mean period. The overall correlation coefficient is 0.1.   The holes indicate that the measured TDIC can not pass the $t$-test.}\label{fig:13hourTDIC}
\end{figure}

\begin{figure} [!htb]
\centering
  \includegraphics[width=0.7\textwidth,clip]{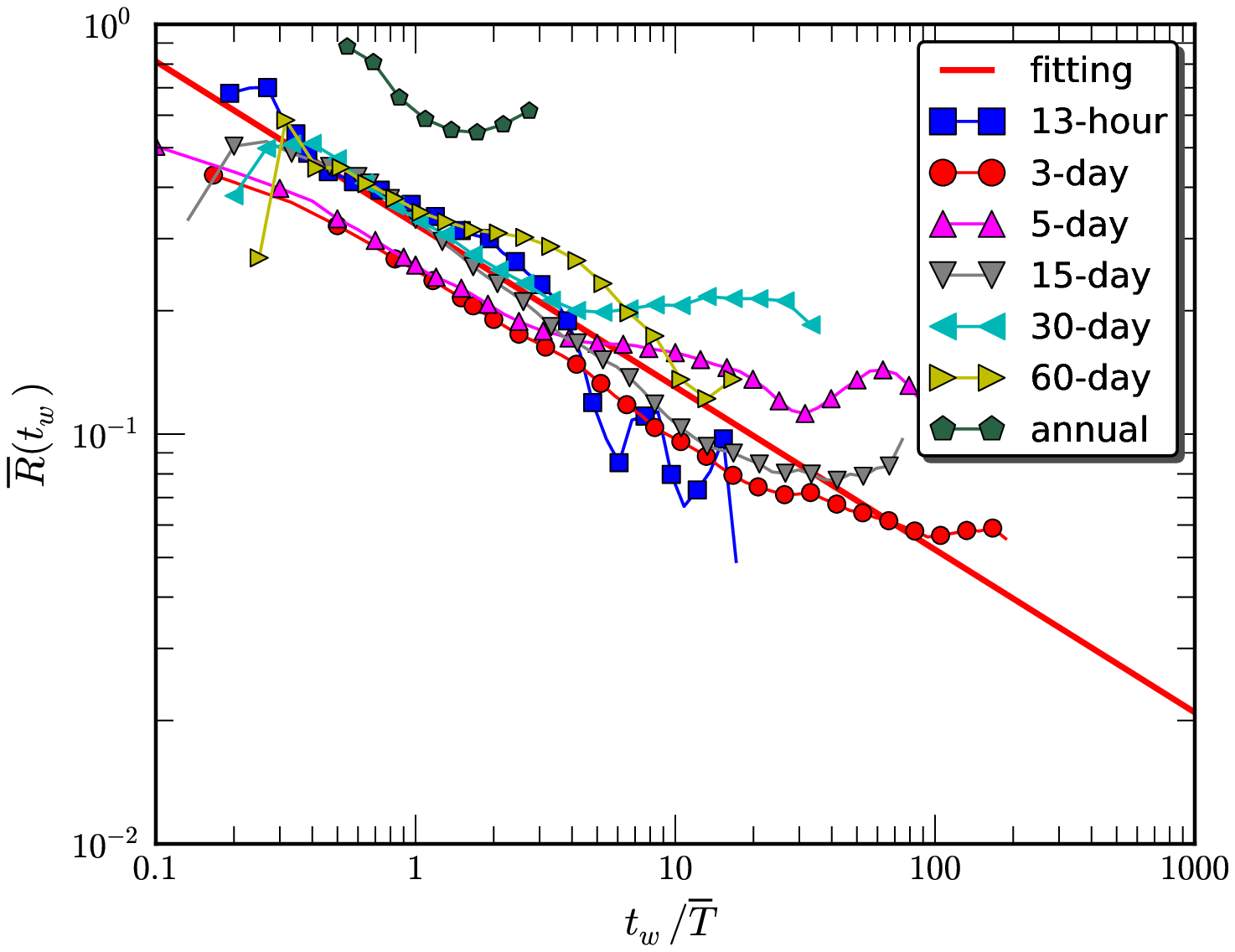}
\caption{ The measured absolute mean value $\overline{R}(t_w)$ for different time scales $\overline{T}$. For display clarity, the sliding window $t_w$ is normalized by  their owner mean period $\overline{T}$. Power law behavior is observed on the  $0.4<t_w/\overline{T}<4$ with a scaling exponent $0.40\pm0.01$.}\label{fig:mTDIC}
\end{figure}

The correlation between this two data sets is considered here. 
Figure \ref{fig:globalcorrelation} displays the global cross 
correlation $R(\tau)$ between the dissolved oxygen and the 
temperature.  There is strong annual cycle with a 30 
days phase difference.  The global correlation coefficient is 
found to be $R(0)=-0.37\pm0.10$. To show the variation of 
$R$, the cross correlation coefficient 
$R(0,t_w)$ is calculated with a sliding window $t_w=180$ days, which is 
shown in Fig.\ref{fig:globalcorrelation} b. The standard 
deviation  is found to be $0.40$. Note that  the standard deviation as a 
measurement of the stationarity has been introduced, see Eq.\eqref{eq:dos}. It 
confirms that the global cross correlation coefficient ignores the local/multi-scale information, which could be recovered by a proper methodology \citep{Chen2010AADA}.

After the EMD decomposition, the data sets are represented 
in a multi-scale way 
\citep{Huang1998EMD,Flandrin2004EMDa,Huang2009PHD}. 
These are used for the multi-scale correlation analysis.  Time scales larger than or equal to one year are now considered.
Figure \ref{fig:trend}  shows the residual from the EMD algorithm, which has been recognized  as the trend of the given data \citep{Wu2007,Moghtaderi2011trend}. 
 Figure \ref{fig:3year}\,a) shows the IMF modes with a 3-year mean period, and Figure \ref{fig:3year}\,b) the corresponding cross correlation $R(\tau)$.  They show a  negative correlation with each other with a phase difference of 2 months.
Figure \ref{fig:annual}\,a) shows the annual cycle from EMD, 
 Figure \ref{fig:annual}\,b) the corresponding cross correlation $R(\tau)$. Again they are negatively 
correlated with each other.  The overall correlation coefficient is -0.63 with a 14 
\red{day} phase difference.  An interesting observation is that there is a weak period 
between 2007 and 2009 for both data sets, indicating that a special event has  happened. The measured TDIC is shown in Figure
\ref{fig:annualTDIC}. There is a positive correlation around 2008. The transition 
from negative (resp. out-of-phase) to positive (in-phase) and again to negative 
correlation from  mid of 2007 to the mid of 2008 might be an effect of the warming from  mid 2007  until almost 2009.  During this period, the sea water temperature was \red{consequently} $2\,^{\circ}\mathrm{C}$ higher than usual in the winter time. Therefore, the out-of-phase relation between DO and temperature is affected by this event.      

Let us consider scales \red{less} than one year.  Figure \ref{fig:60day} shows the IMF modes with a mean period \red{of} 60 days.  They are positively correlated with each other 
on some portions and negatively correlated on \red{others}, showing rich dynamics. However, the overall correlation coefficient is small (resp. $-0.02$).  Figure 
\ref{fig:60dayTDIC} displays the  measured TDIC, which confirms the direct observation of the IMF modes. A strong positive 
correlation is observed \red{during} the time period 2009$\sim$2010.  
Due to the complexity of the coastal environment system, the exact reason for 
this changes from negative to positive is unclear. It will be  addressed in future  
studies.
For  time scales smaller than 60 
days, the original IMF are not shown here, but only the measured TDIC.  Figure 
\ref{fig:30dayTDIC}$\sim$\ref{fig:13hourTDIC} 
 show the measured TDIC  for mean 
periods of 30-day to 13-hour respectively. 
All TDICs show rich \red{patterns} at \red{a} small sliding window.
Let us  also note that with the increase of the window size they \red{ decorrelate}. 
To show the decorrelation of the TDIC, a absolute mean value of the measured $R_i(t_k^n)$ is defined, i.e.,
\begin{equation}
\overline{R}(t_w)=\langle | R(t_k^n,t_w)  |  \rangle_t 
\end{equation}
Figure \ref{fig:mTDIC} shows  in a log-log plot of  the measured $\overline{R}(t_w)$ 
for seven time scales considered above.  Note that the horizontal axis $t_w$ has been normalized by their mean period $\overline{T}$. Generally speaking, the curves are decreasing 
with the increasing of the sliding window size 
$t_w/\overline{T}$, showing the decorrelation between 
dissolved oxygen and temperature on different scales. A power law behavior is observed on $0.4<t_w/\overline{T}<4$, i.e.,
\begin{equation}
\overline{R}\sim (t_w/\overline{T})^{-\alpha}
\end{equation}
with a scaling exponent $\alpha=0.40\pm0.10$ obtained  by using a least square fitting algorithm. This indicates that the considered modes  decorrelate with each other in a power law way, with the same law.

\section{Conclusion}\label{sec:conclusion}
Marine environmental time series can be quite complex, \red{especially} with \red{the} influence of turbulent stochastic fluctuations, multi-scale dynamics, \red{and} deterministic forcings. Such time series are \red{as a result} often non stationary and nonlinear. It has been shown elsewhere that the EMD method, together with Hilbert spectral analysis, is less influenced
by deterministic forcing than other methods \citep{Huang2010PRE,Huang2011PRE}. Here this approach was applied to automatic temperature and oxygen data. 
Power law spectra was found using Hilbert spectral analysis.
Both series have similar spectra over the range from 2 to 100 days, whereas there is a difference in slopes for
high frequency power law regimes.
The time evolution and scale dependence of cross correlations between both series were considered. The decomposition into modes helped to estimate how correlations vary among scales.
It was found that the trends are perfectly anti-correlated, and that the modes of mean year 3 years and 1 year have also negative correlation, whereas higher frequency modes have a much smaller correlation. A new methodology
was also applied,  time-dependent
intrinsic correlations: this showed the patterns of correlations at different scales, for different modes.

Such analysis may help to identify some key mechanisms. Here a strong anti-correlation was found for
modes of mean time scale of 3 years, with a phase shift of 2 months. An explanation of such \red{a} result could be
that, at these scales, higher temperatures may favour larger phytoplankton growth rate, and hence, with a time
delay, \red{a} lower percentage of oxygen. Of course such relations are not always true, and the color maps of TDIC reveal the variations of the strength of such relationship. 

While Fourier space methods can produce in some respects similar results for the power spectra, cross-correlations through co-spectra, and some colour maps of correlations in frequency space, they do not provide time-frequency information like the EMD based method used here. Furthermore, the TDIC tested here 
is a new method for correlation analysis that \red{could} be applied on other time series from the environmental and oceanic sciences, since in these fields time series are typically complex, with fluctuations at all scales, and nonlinear and non stationary features.

\noindent\section*{Acknowledgements}
This work is sponsored in part  by the National Natural Science
Foundation of China (Nos. 11072139, 11032007 and  11202122), and \lq{}Pu Jiang\rq{} project of Shanghai (No. 12PJ1403500) and  the Shanghai Program for Innovative Research Team in Universities.   The EMD \textsc{Matlab} codes used in this study are
written by Dr Gabriel Rilling and Prof. Patrick Flandrin from \red{the} laboratoire de Physique, CNRS \& ENS
Lyon (France): http://perso.ens-lyon.fr/patrick.flandrin/emd.html. We thank Ifremer, and especially Alain Lefebvre and Michel Repecaud for their work in recording and validating the Marel data. We thank the I. Puillat and the others organizers of the Time Series Brest conference for a very nice meeting, and the reviewers for useful comments and suggestions. We thank also Denis Marin (LOG, Dunkerque, France) for the realization of Figure 1. 

\section*{References}
\bibliographystyle{elsarticle-harv}

\end{document}